\documentclass[doublecol]{epl2c}

\usepackage{graphicx}
\usepackage{dcolumn}
\usepackage{amsmath}
\usepackage{color}

\title{Phase diagram of the three-band d-p model based on the
optimization variational Monte Carlo method
}

\shorttitle{Phase diagram of the three-band d-p model}

\author{$^a$Takashi Yanagisawa, $^b$Mitake Miyazaki,
$^a$Kunihiko Yamaji
}

\institute{$^a$Electronics and Photonics Research Institute,
National Institute of Advanced Industrial Science and Technology (AIST),
Tsukuba Central 2, 1-1-1 Umezono, Tsukuba 305-8568, Japan\\
$^b$Hakodate Institute of Technology, 14-1 Tokura, Hakodate,
Hokkaido 042-8501, Japan
}

\pacs{71.10.-w}{First pacs description}
\pacs{71.27.+a}{Second pacs description}
\pacs{71.10.Fd}{Third pacs description}

\abstract{
The phase diagram of cuprate high-temperature superconductors is
investigated on the basis of the three-band d-p model.
We use the optimization variational Monte Carlo method, where improved
many-body wave functions have been proposed to make the ground-state
wave function more precise. 
We investigate the stability of antiferromagnetic state by changing
the band parameters such as the hole number, level difference $\Delta_{dp}$ between
$d$ and $p$ electrons and transfer integrals.
We show that the antiferromagnetic correlation weakens when $\Delta_{dp}$
decreases and the pure $d$-wave superconducting phase may exist in this region.
We present phase diagrams including antiferromagnetic and superconducting
regions by varying the band parameters. 
The phase diagram obtained by changing the doping rate $x$ contains 
antiferromagnetic, superconducting and also phase-separated phases. 
We propose that high-temperature superconductivity will occur 
near the antiferromagnetic boundary in the space of band parameters.
}

\begin{document}
\maketitle

\section{Introduction}

The physics of high-temperature superconductivity has been studied
from the viewpoint of strongly correlated electron systems since its
discovery\cite{bed86}.
This is because cuprate parent materials are Mott insulators and the
Cooper pairs have the d-wave symmetry.
The fundamental model of cuprate  high-temperature superconductors is
the electronic model for the CuO$_2$ plane including copper and oxygen
atoms.
This model is called the d-p model (or three-band Hubbard
model)\cite{eme87,hir89,sca91,ogu94,koi00,yan01,koi01,yan03,koi03,koi06,yan09,web09,lau11,web14,ave13,ebr16,tam16}.
It is certain that the electron correlation plays an important role in the d-p model
since the on-site Coulomb interaction $U_d$ between d electrons is large.
This makes it a difficult task to understand the phase diagram of the d-p model.
Because $U_d$ is the large-energy scale interaction, new phenomena
including high-temperature superconductivity can be expected.
Simplified models such as the two-dimensional single-band 
Hubbard 
model\cite{hub63,hub64,gut63,zha97,zha97b,yan96,nak97,yam98,yam11,har09,bul02,yok04,yok06,aim07,miy04,yan08,yan13,yan16,yan19,yan19c,yan21,yan21a}
or ladder model\cite{noa95,noa97,yam94,koi99,yan95,nak07} have been studied to 
investigate whether the attractive interaction is induced from the
on-site Coulomb interaction.

We employ the optimization variational Monte Carlo method to investigate
electronic properties of the three-band d-p model.
In this method we calculate the expectation values of several physical
properties numerically using a Monte Carlo algorithm\cite{yan98,bla81,yan07}.
In order to control the strong electron correlation, we use the
Gutzwiller function and wave functions that are improved
by multiplying an initial wave function by $\exp(-K)$-type 
operators\cite{yan16,yan98}, where $K$ is a correlation operator with
variational parameters. In this paper, $K$ stands for the non-interacting
part of the Hamiltonian.
In the single-band Hubbard model, the ground-state energy is lowered
greatly and becomes lower than that obtained by other wave functions\cite{yan16}.

The single-band Hubbard model is the fundamental model and was used
to understand the metal-insulator transition\cite{mott} and magnetic
properties\cite{yos96}.
It has been shown from numerical studies based on improved wave functions
that the superconducting phase exists in the
single-band two-dimensional (2D) Hubbard model\cite{yok06,yan16,yan19,mis14,yan99}. 
We regard the multi-band model as a more realistic
model of high-temperature superconductors.

In this paper we investigate the stability of antiferromagnetically ordered
state in the two-dimensional d-p model.  The phase diagram is determined
as a result of the competition between antiferromagnetic (AF) and 
superconducting (SC) pair correlations. 
The AF region decreases as the level difference $\epsilon_p-\epsilon_d$
decreases and thus the pure d-wave pairing phase may exist when 
$\epsilon_p-\epsilon_d$ is small.
There is a phase separation (PS) in the low-doping region as in the 
two-dimensional Hubbard model\cite{can91,cap95,lor02,gun07,toc16}.  
We show the phase diagram as a function
of the doping rate to show AF, SC and PS regions in the d-p model.
We think that high-temperature superconductivity occurs in the crossover
region near the boundary of AF phase\cite{yan16,yan19}.
In the single-band Hubbard model, the crossover takes place as the on-site
Coulomb interaction $U$ increases (or decreases).
In the three-band d-p model, there are several parameters that can
control AF correlation and would induce crossovers where AF or charge
fluctuations grow larger.


\section{Hamiltonian}

We consider the three-band d-p model in this paper.
The Hamiltonian is written as
\begin{align}
H&= \epsilon_d\sum_{i\sigma}d_{i\sigma}^{\dag}d_{i\sigma}
+ \epsilon_p\sum_{i\sigma}(p_{i+\hat{x}/2\sigma}^{\dag}p_{i+\hat{x}/2\sigma}
\nonumber\\
&+ p_{i+\hat{y}/2\sigma}^{\dag}p_{i+\hat{y}/2\sigma})
\nonumber\\
&+ t_{dp}\sum_{i\sigma}[d_{i\sigma}^{\dag}(p_{i+\hat{x}/2\sigma}
+p_{i+\hat{y}/2\sigma}-p_{i-\hat{x}/2\sigma}-p_{i-\hat{y}/2\sigma})
\nonumber\\
&+ {\rm h.c.}]\nonumber\\
&+ t_{pp}\sum_{i\sigma}[p_{i+\hat{y}/2\sigma}^{\dag}p_{i+\hat{x}/2\sigma}
-p_{i+\hat{y}/2\sigma}^{\dag}p_{i-\hat{x}/2\sigma}\nonumber\\
&- p_{i-\hat{y}/2\sigma}^{\dag}p_{i+\hat{x}/2\sigma}
+p_{i-\hat{y}/2\sigma}^{\dag}p_{i-\hat{x}/2\sigma}+{\rm h.c.}]\nonumber\\
&+ t_d'\sum_{\langle\langle ij\rangle\rangle\sigma}
(d_{i\sigma}^{\dag}d_{j\sigma}
+{\rm h.c.} )
+ U_d\sum_i d_{i\uparrow}^{\dag}d_{i\uparrow}d_{i\downarrow}^{\dag}
d_{i\downarrow}.
\end{align}
$d_{i\sigma}$ and $d^{\dag}_{i\sigma}$ represent the operators for the $d$ hole,
and $p_{i\pm\hat{x}/2\sigma}$ and $p^{\dag}_{i\pm\hat{x}/2\sigma}$ denote the
operators for the $p$ holes at the site $R_{i\pm\hat{x}/2}$, and in a
similar way $p_{i\pm\hat{y}/2\sigma}$ and $p^{\dag}_{i\pm\hat{y}/2\sigma}$
are defined.
$t_{dp}$ and $t_{pp}$ represent the transfer integrals between adjacent Cu and O 
orbitals and that between nearest neighbor p orbitals, respectively.
$t_d'$ was introduced to reproduce the Fermi 
surface\cite{yan14} for cuprate superconductors. 
$U_d$ denotes the on-site Coulomb repulsion and $U_p$ is that between
$p$ holes.
In general, $U_p$ is smaller than $U_d$\cite{web08,hyb89,esk89,mcm90,esk91}.

The values of band parameters have been estimated in the literature. 
For example, $U_d=10.5$, $U_p=4.0$
and $U_{dp}=1.2$ in eV\cite{hyb89} where $U_{dp}$ is the nearest-neighbor
 Coulomb interaction
between holes on adjacent Cu and O orbitals.
We neglect $U_{dp}$ because $U_{dp}$ is small compared to $U_d$.
The level difference between $d$ and $p$ holes is denoted as 
$\Delta_{dp}=\epsilon_p-\epsilon_d$.
$N$ stands for the number of sites and the total number of atoms is
given by $N_a=3N$.
The energy measured in units of $t_{dp}$ in this paper.
We use the hole picture and we consider the case where $\Delta_{dp}\geq 0$
in this paper.


\begin{figure}[ht]
\begin{center}
\includegraphics[width=7.0cm]{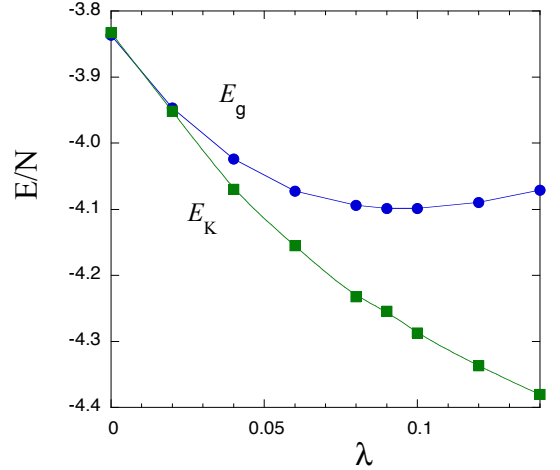}
\caption{\label{ek76}
Ground-state energy $E_g$ and the kinetic energy $E_K=\langle K\rangle$ 
as a function of $\lambda$ on an $8\times 8$ lattice with 76 holes.
The band parameters are $\epsilon_p-\epsilon_d=2t_{dp}$,
$t_{pp}=0.4$, $t_d'=-0.2$, $U_d=10$ and $U_p=0$ where the energy unit
is $t_{dp}$.
}
\end{center}
\end{figure}

\begin{figure}[ht]
\begin{center}
\includegraphics[width=8.0cm]{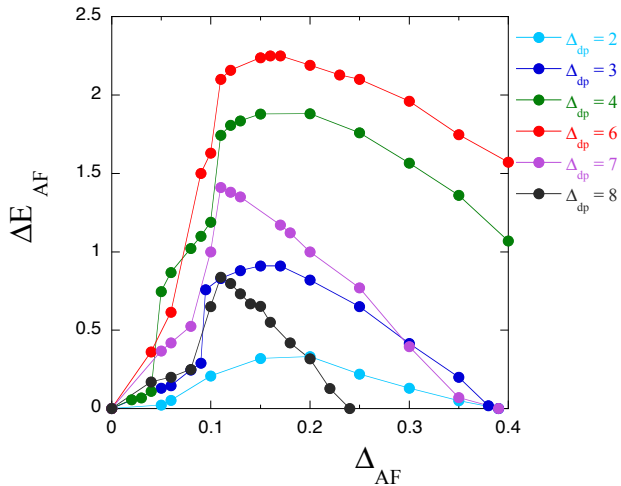}
\caption{\label{fig2}
Antiferromagnetic condensation energy for $\psi_{\lambda}$
as a function of the AF
gap parameter $\Delta_{AF}$ for several values of the level difference
$\Delta_{dp}\equiv\epsilon_{p}-\epsilon_d$.
We carried out calculations on an $8\times 8$ lattice (192 atoms)
with 76 holes (where the doping rate $x$ is $x=0.1875$).
We set $t_{pp}=0.4$, $t_d'=0$ and $U_d=10$ in units of $t_{dp}$.
}
\end{center}
\end{figure}

\section{Improved wave functions with electron correlation}

The wave function is given as
\begin{equation}
\psi_{\lambda}= \exp(-\lambda K)\psi_G,
\end{equation}
where $K$ denotes the non-interacting part of the Hamiltonian:
$K=H$ with $U_d=0$.
$\lambda$ is a real variational parameter.
$\psi_G$ is given as 
$\psi_G = P_G\psi_0$ ,
with $P_G= \prod_i[1-(1-g)n_{di\uparrow}n_{di\downarrow}]$ where
$g$ takes the value in the range $0\leq g\leq 1$. 
This is the wave function which is improved from the Gutzwiller
function\cite{yan16,yan98,ots92,yan99,eic07,bae09,bae11,bae19,yan19}
generalized to the three-band model\cite{yan19d} by multiplying by 
an exponential factor.
$\psi_0$ is a one-particle state with variational parameters
$\tilde{t}_{dp}$, $\tilde{t}_{pp}$, $\tilde{t}'_d$,
$\tilde{\epsilon}_p-\tilde{\epsilon}_d$\cite{yan19d}:
$\psi_0= \psi_0(\tilde{t}_{dp}, \tilde{t}_{pp}, \tilde{t}'_d,
\tilde{\epsilon}_p-\tilde{\epsilon}_d)$.
$\psi_0$ is obtained as the eigenstate of the non-interacting
Hamiltonian with these variational parameters.
We set $\tilde{t}_{dp}=t_{dp}$ and the energy is measured in units
of $t_{dp}$.
$K$ also contains variational parameters:
$K= K(\tilde{t}_{pp}, \tilde{t}'_d,
\tilde{\epsilon}_p-\tilde{\epsilon}_d)$.
The expectation values are calculated by using the variational
Monte Carlo method.

We here give a discussion on the role $K$ in the wave function.
The operator $e^{-\lambda K}$ controls the weights of excitation
modes in the Gutzwiller function.  $e^{-\lambda K}$ suppresses
high-energy excitations for which the eigenvalues of $K$ are large
and thus $e^{-\lambda k}$ becomes small.  This indicates that
$e^{-\lambda K}$ plays a role of projection that projects out low 
lying excitation modes from the wave function.
This appears in the behavior of the momentum distribution
function 
$n_{{\bf k}\sigma}\equiv \langle c_{{\bf k}\sigma}^{\dag}c_{{\bf k}\sigma}\rangle$.
In fact, $n_{{\bf k}\sigma}$ evaluated by the Gutzwiller function
shows an unphysical behavior, that is, $n_{{\bf k}\sigma}$ at 
${{\bf k}}$ near the Fermi surface is often larger than that at
the bottom of the band\cite{yok86}.  This shortcoming is remedied by the
improved wave function\cite{yan14}. 

We show the energy-expectation value as a function of $\lambda$ in
Fig. 1 where calculations were
performed on the system of $8\times 8$ lattice with 76 holes.
$E_K$ denotes the expectation value $E_K=\langle K\rangle$.
The parameters are $\Delta_{dp}=\epsilon_p-\epsilon_d=2$, $t_{pp}=0.4$,
$t_d'=-0.2$ and $U_d=10$ in units of $t_{dp}$.
The lowering of the total energy originates from the kinetic-energy
gain as $\lambda$ is increased.
The total energy given by the sum of the kinetic energy and the
potential energy has a minimum at finite value of $\lambda$.
Since the Coulomb potential energy increases as $\lambda$ increases,
the total energy is determined by the balance between
the kinetic energy and the Coulomb energy.

The correlated BCS wave function is written as
\begin{equation}
\psi_{\lambda-BCS} = 
= \exp(-\lambda K)P_G\prod_{\bf k}(u_{\bf k}\beta_{\bf k}^{\dag}
+v_{\bf k}\alpha_{\bf k}^{\dag})|\tilde{0}\rangle,
\label{pbcsfn2}
\end{equation}
where the particle-hole transformation for down-spin holes\cite{yan16,yan99}
has been performed: 
$\beta_{\bf k}^{\dag}=\alpha_{-{\bf k}\downarrow}$ and
$\alpha_{\bf k}^{\dag}=\alpha_{{\bf k}\uparrow}^{\dag}$.
$|\tilde{0}\rangle$ denotes the vacuum for newly defined $\alpha$
and $\beta$ particles satisfying 
$\alpha_{\bf k}|\tilde{0}\rangle=\beta_{\bf k}|\tilde{0}\rangle=0$.
The BCS parameter is given by the conventional form
$v_{\bf k}/u_{\bf k}=\Delta_{\bf k}/(\xi_{\bf k}+\sqrt{\xi_{\bf k}^2+\Delta_{\bf k}^2})$,
where $\xi_{\bf k}$ is the dispersion relation of the lowest band
we assume the d-wave symmetry $\Delta_{\bf k}=\Delta(\cos k_x-\cos k_y)$.
$\Delta=\Delta_{sc}$ is a variational parameter and the optimized value
$\Delta_{sc}$ is regarded as the superconducting gap.
In this representation, the electron pair operator
$\alpha_{{\bf k}\uparrow}^{\dag}\alpha_{{\bf k}\downarrow}^{\dag}$ 
is transformed to the hybridization form
$\alpha_{{\bf k}}^{\dag}\beta_{{\bf k}}$.
The chemical potential is used to adjust the expectation value
of the total electron number.


\section{Antiferromagnetic state and the level difference}

The antiferromagnetic wave function is formulated by introducing
the AF order parameter $\Delta_{AF}$ in the initial wave function
$\psi_0(\Delta_{AF})$:
\begin{equation}
\psi_{\lambda}= \exp(-\lambda K)P_G\psi_0(\Delta_{AF}).
\end{equation}
The level difference $\Delta_{dp}=\epsilon_p-\epsilon_d$ plays an
important role in determining the stability of the AF state.
We show the AF condensation energy $\Delta E_{AF}$ as a function of the AF
order parameter $\Delta_{AF}$ for several values of the
level difference between d and p electrons in Fig. 2.
We carried out calculations on a system of $8\times 8$ lattice
with 192 atoms including copper and oxygen atoms.
$\Delta E_{AF}$ has a maximum when the level difference $\Delta_{dp}$
is about half of $U_d$, namely, near the 'symmetric case' with
$\Delta_{dp}=U_d/2$.
In Fig. 2, we set the parameters as $t_{pp}=0.4$, $t_d'=0$ and $U_d=10$
in units of $t_{dp}$, and the doping rate is $x=0.1875$.
We show $\Delta E_{AF}/N$ as a function of the level difference
$\Delta_{dp}$ for $t_d'=0$ and $-0.2$ in Fig. 3.
The AF order parameter $\Delta_{AF}$ at which $\Delta E_{AF}$ takes
a maximum value is shown in Fig. 4 for each value of $\Delta_{dp}$.
For $t_d'=0$, $\Delta_{AF}$ vanishes when $\Delta_{dp}\simeq 1$. 
This indicates that we may have a superconducting phase when the d-hole
level is near the p-hole level.
The results in Figs. 3 and 4 also show that $t_d'$ suppresses the
magnetic instability.

We show $\Delta E_{AF}/N$ as a function of $\Delta_{dp}$ when the
doping rate is $x=0.125$ in Fig. 5.
The Fig. 5 shows that $\Delta E_{AF}$ has also a maximum for $\Delta_{dp}\sim U_d/2$
and AF correlation survives even up to the region of small $\Delta_{dp}\sim 1$ 
for small $x$.

\begin{figure}[ht]
\begin{center}
\includegraphics[width=7.5cm]{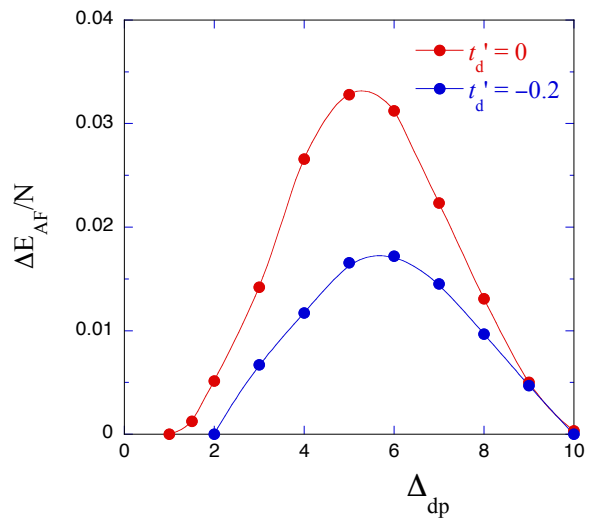}
\caption{\label{fig3}
Antiferromagnetic condensation energy for $\psi_{\lambda}$
as function of the level difference
$\Delta_{dp}= \epsilon_{p}-\epsilon_d$ on an $8\times 8$ lattice with
76 holes.
We used $t_{pp}=0.4$ and $U_d=10$  in units of $t_{dp}$.
}
\end{center}
\end{figure}


\begin{figure}[ht]
\begin{center}
\includegraphics[width=7.5cm]{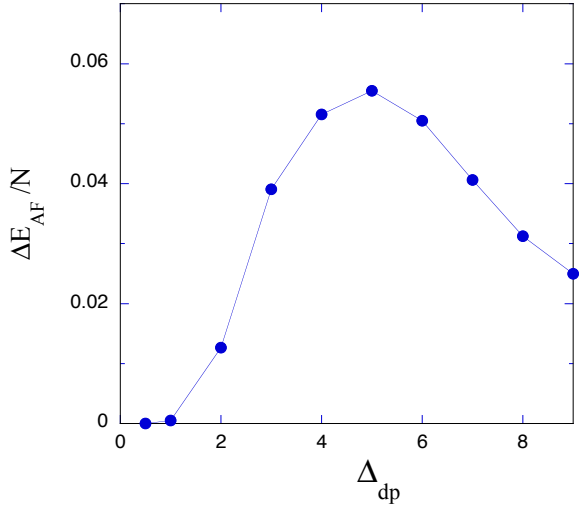}
\caption{\label{fig4}
Antiferromagnetic condensation energy for $\psi_{\lambda}$
as function of the level difference
$\Delta_{dp}$ on an $8\times 8$ lattice with 72 holes ($x=0.125$).
We set $t_{pp}=0.4$, $t_d'=0$ and $U_d=10$.
}
\end{center}
\end{figure}

\begin{figure}[ht]
\begin{center}
\includegraphics[width=7.5cm]{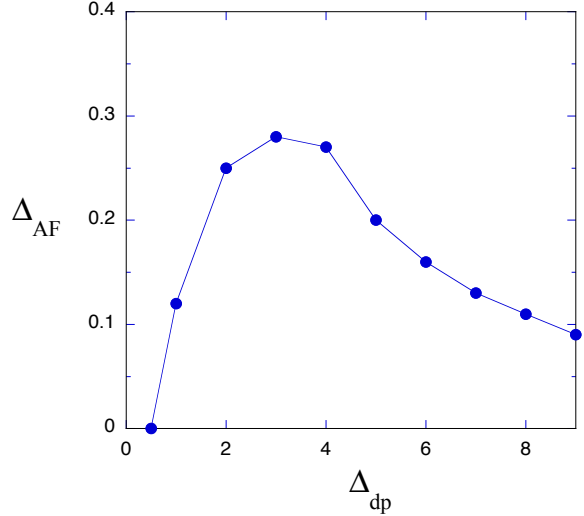}
\caption{\label{fig5}
Antiferromagnetic order parameter for $\psi_{\lambda}$
as function of the level difference
$\Delta_{dp}$ on an $8\times 8$ lattice with 72 holes.
The parameers are the same as in Fig.5.
}
\end{center}
\end{figure}

\section{Superconducting state}

First we examine the superconducting state by employing the
BCS-Gutzwiller wave function 
$\psi_{G-BCS} = P_{N_e}P_G\psi_{BCS}$,
where $\psi_{BCS}$ indicates the BCS wave function.
$P_{N_e}$ is the number-projection operator that extracts only the
states with a fixed total hole number.
We show the SC condensation energy as a function of the SC
order parameter $\Delta_{sc}$ in Fig.6 for several values of the
level difference $\Delta_{dp}$.  The result shows that the SC condensation
energy increases as $\Delta_{dp}$ increases.
Based on the BCS-Gutzwiller function, the large level difference
is more favorable for superconductivity.
We should, however, remember instability toward antiferromagnetic state
when the level difference $\Delta_{dp}$ becomes large.


From the competition between AF and SC orderings,
we expect that the pure $d$-wave SC state is stabilized when
$\epsilon_d-\epsilon_p$ is near the boundary of the
AF region.  
The AF and SC order parameters are shown as a function of $\Delta_{dp}$
in Fig. 7 for $x=0.1875$.
We included the SC order parameter for the Gutzwiller function
for reference when $\Delta_{dp}\geq 2t_{dp}$ although SC state
is no longer stable in this region.
This figure indicates that the SC state
exists when the level difference is small and that
high-temperature superconductivity will occur in the small 
$\Delta_{dp}$-region.

\begin{figure}[ht]
\begin{center}
\includegraphics[width=7.5cm]{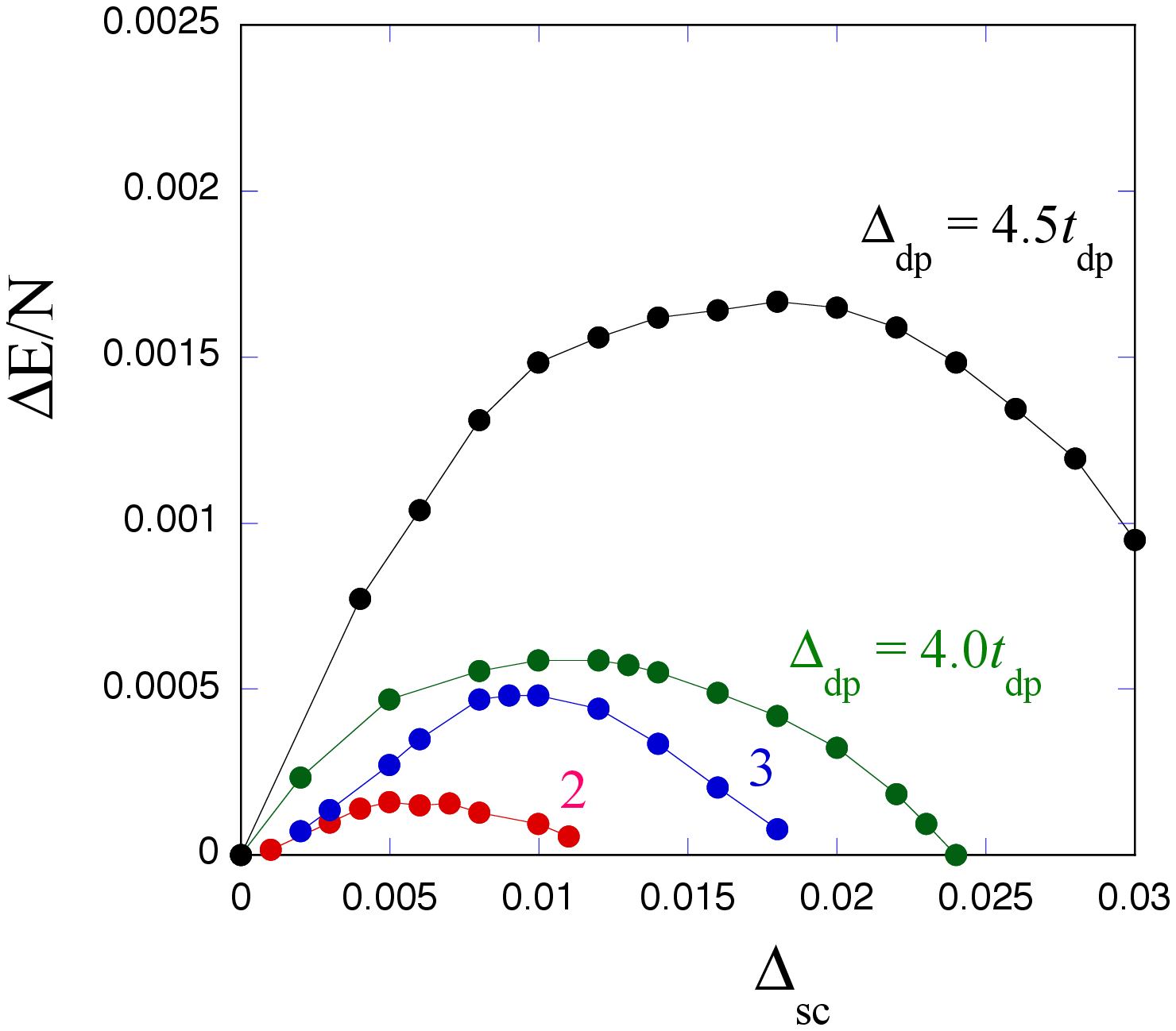}
\caption{\label{fig6}
SC condensation energy for the BCS-Gutzwiller function
as function of the superconducting
gap parameter $\Delta_{sc}$ for several values of the level difference
$\Delta_{dp}\equiv\epsilon_{p}-\epsilon_d$.
The calculations were carried out on an $8\times 8$ lattice with
$t_{pp}=0.4$, $t_d'=0$ and $U_d=10$ in units of $t_{dp}$.
The number of holes is 76 for $\Delta_{dp}=3$, 4 and 4.5 and 72
for $\Delta_{dp}=2$ (closed shell cases).
}
\end{center}
\end{figure}


\begin{figure}[ht]
\begin{center}
\includegraphics[width=7.5cm]{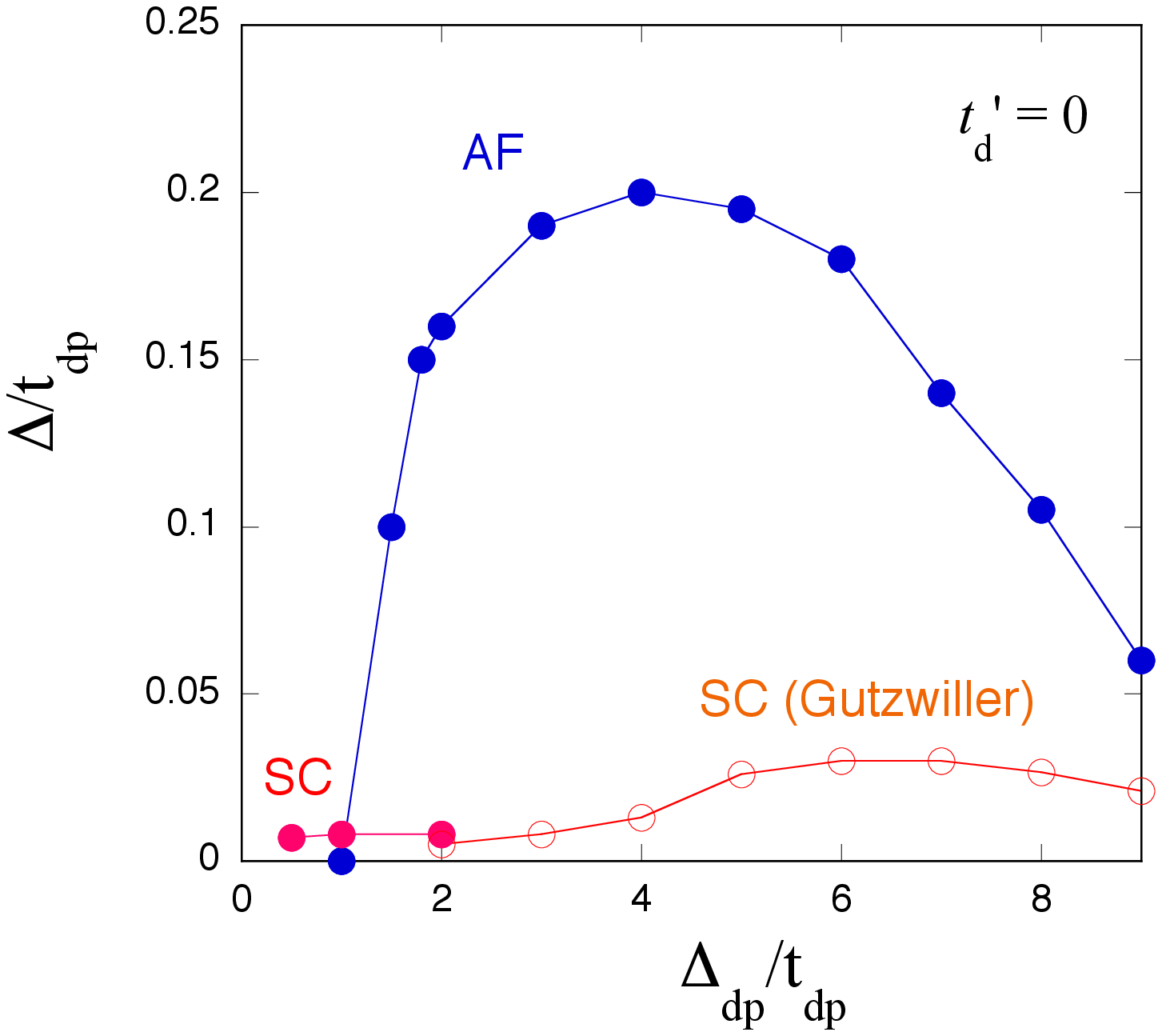}
\caption{\label{fig7}
AF and SC order parameters for the $\psi_{\lambda}$ wave function 
as a function of the level difference $\Delta_{dp}$.
The SC order parameter obtained by the Gutzwiller function is
also shown by open circles for reference in the region 
$\Delta_{dp}\geq 2t_{dp}$. 
We set $x=0.1875$ and the band parameters are $t_{pp}=0.4$,
$t_d'=0$ and $U_d=10$.
}
\end{center}
\end{figure}

\begin{figure}[ht]
\begin{center}
\includegraphics[width=7.5cm]{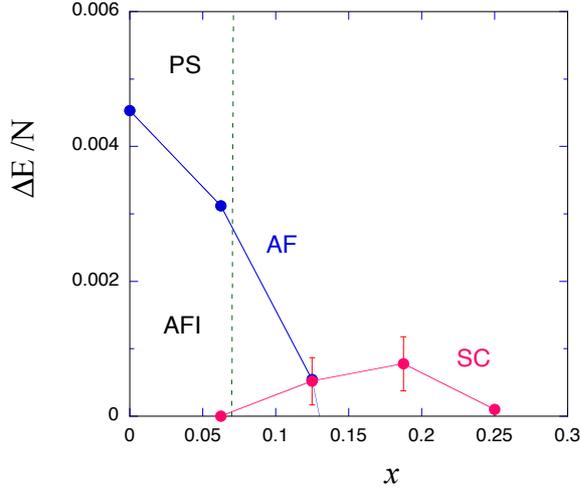}
\caption{\label{fig8}
AF and SC condensation energies as a function of the hole doping rate $x$
for $\Delta_{dp}= 1$ on a $8\times 8$ lattice with 
$t_{pp}=0.4$, $t_d'=0$ and $U_d=10$ in units of $t_{dp}$.
There is a phase-separated region when $x$ is small.
}
\end{center}
\end{figure}

\begin{figure}[ht]
\begin{center}
\includegraphics[width=7.5cm]{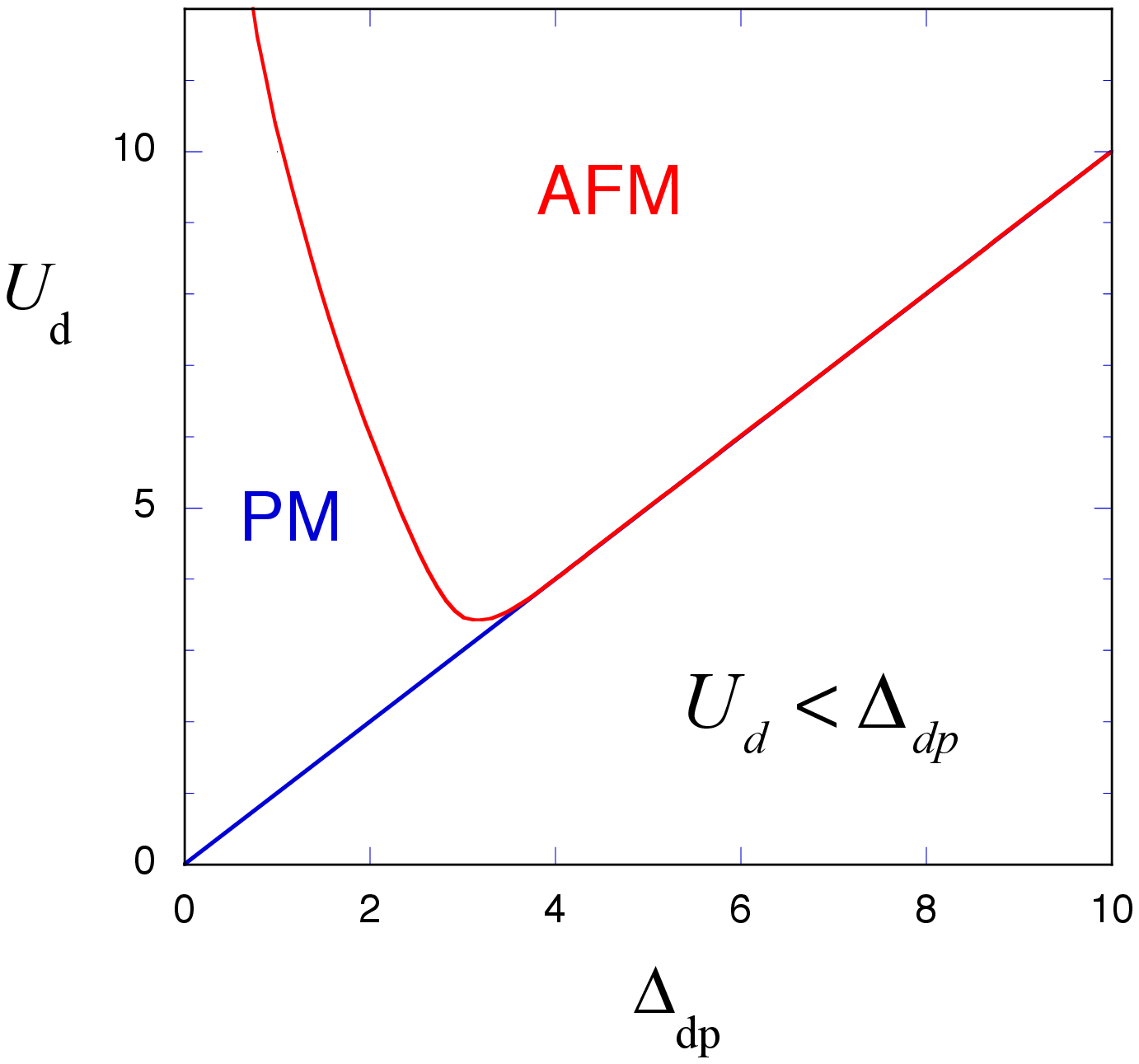}
\caption{\label{fig9}
AF region in the plane of $U_d$ and the level difference
$\Delta_{dp}$ for $x=0.1875$ on an $8\times 8$ lattice.
We set $t_{pp}=0.4$ and $t_d'=0$, and the energy is measured in units
of $t_{dp}$.
}
\end{center}
\end{figure}

\section{Phase diagram and phase separation}

Let us examine the possible phase diagram of cuprate superconductors
based on the optimization variational Monte Carlo method.
We exhibit the condensation energy as a function of the doping rate $x$
for $\Delta_{dp}=t_{dp}$ in Fig. 8. 
There is the AF region when $x$ is small and the SC region
exists near the optimum region $x\sim 0.2$.
We should mention that there is a phase-separated region in the
low-doping region where $x< 0.07$.
This indicates the existence of AF insulator phase in the low doping region.
This is similar to the phase diagram of the 2D Hubbard model.
The phase separation (PS) is, however, dependent on the level difference $\Delta_{dp}$.
As $\Delta_{dp}$ decreases, the phase-separated region decreases and
vanishes when $\Delta_{dp}$ approaches zero.
Thus the area of phase-separated region can be controlled by changing
the band parameters.

We lastly show the phase diagram in the $U_d$-$\Delta_{dp}$ plane
for $x=0.1875$ with $t_{pp}=0.4t_{dp}$ and $t_d'=0$ in Fig. 9.
There are a large AF region and paramagnetic(PM) phase in Fig. 9
where SC regions are not explicitly included.  We expect that
high-temperature superconductivity may be realized in the relatively
small region near the boundary between AF and PM regions.

\section{Summary}

We have investigated the ground-state properties of the three-band
d-p model on the basis of the optimization variational Monte Carlo method.
The ground-state energy is lowered greatly by multiplying by the
exponential operator $e^{-\lambda K}$ as in the 2D Hubbard model.
In general, the antiferromagnetic correlation is suppressed as the
wave function is improved.
  
In the d-p model the AF correlation is very much stronger
than that in the single-band Hubbard model.  It is thus important to
control the strength of AF correlation by varying the interaction and
band parameters.  The level difference $\Delta_{dp}$ plays an important
role in the study of the phase diagram of the d-p model.
The pure $d$-wave superconducting phase exists when $\Delta_{dp}$ becomes small.
In particular high-temperature superconductivity is expected near the
AF boundary in the small $\Delta_{dp}$ region. 

A phase-separated region exists in the d-p model as in the single-band
2D Hubbard model.  In this region the ground state is an insulator with
AF order and is dependent upon $\Delta_{dp}$.  
The PS region increases as $\Delta_{dp}$ increases and decreases and
vice versa.
The close relation between PS region and incommensurate phases have
been examined\cite{lor02,gun07}.
The PS instability may be related to an instability toward some charge order
such as stripes.

The phase diagram in Fig. 9 is consistent with several experiments
concerning the critical temperature $T_c$ of cuprate 
superconductors\cite{zhe95,tak13}.
There is a correlation between $T_c$ and the ratio of hole numbers
$n_p/n_d$ where $n_p$ and $n_d$ denote the hole density of p and
d holes, respectively\cite{zhe95}.  
There is the tendency that $T_c$ is higher for larger $n_p/n_d$,
that is, $T_c$ increases with the decrease of $\epsilon_p-\epsilon_d$.
This is consistent with our result that the pure d-wave pairing
state exists when $\epsilon_p-\epsilon_d$ is small.

The increase of $T_c$ under pressure\cite{tak13} is understood 
within the d-p model
as follows.  $\Delta_{dp}$ and $U_d$ would decrease by applying pressure.
If the first electronic state is not in the optimum SC state near the
AF boundary, $T_c$ will
increase as $\Delta_{dp}$ decreases.

We neglected Cu-O Coulomb repulsion $U_{dp}$ in this paper. 
It has been pointed out that
a small $U_{dp}$ can produce dramatic consequences on the charge
stability\cite{can91,rai93}.
It is a future issue to examine the effect of $U_{dp}$.
It is well known that the parent compounds of high-temperature
cuprates are charge-transfer insulators.
Based on the optimization variational Monte Carlo method, the critical
value of $\Delta_{dp}$ for charge-transfer metal-insulator transition 
is about $2t_{dp}$\cite{yan14}.
In our calculations the favorable value of $\Delta_{dp}$ for
superconductivity is a little bit smaller than this critical value because of
competition between AF and SC correlations.
This would indicate a possibility that the critical temperature $T_c$
can be increased by changing material parameters.
We also mention that the inclusion of $U_{dp}$ or band parameter
$t_d'$ possibly may change the physical picture. 

A part of computations was supported by the Supercomputer Center
of the Institute for Solid State Physics, the University of
Tokyo, and the Supercomputer system Yukawa-21 of the Yukawa Institute
for Theoretical Physics, Kyoto University.
This work was supported by a Grand-in-Aid for Scientific
Research from the Ministry of Education, Culture, Sports,
Science and Technology of Japan (Grant No. 17K05559).

\end{document}